\documentclass[preprintnumbers,amsmath,amssymb,showpacs]{revtex4}
\usepackage{epsfig}
\usepackage{color}
\usepackage[colorlinks,urlcolor=blue,citecolor=red]{hyperref}
\begin{document}
\setlength{\voffset}{1.0cm}
\title{Twisted kink dynamics in multiflavor chiral Gross-Neveu model}
\author{Michael Thies\footnote{michael.thies@gravity.fau.de}}
\affiliation{Institut f\"ur  Theoretische Physik, Universit\"at Erlangen-N\"urnberg, D-91058, Erlangen, Germany}
\date{\today}
\begin{abstract}
The Gross-Neveu model with ${\rm U}_L(N_f)\times{\rm U}_R(N_f)$ chiral symmetry is reconsidered in the large $N_c$ limit. The known
analytical solution for the time dependent interaction of any number of twisted kinks and breathers is cast into a more revealing form.
The ($x,t$)-dependent factors are isolated from constant coefficients and twist matrices. These latter generalize the twist phases 
of the single flavor model. The crucial tool is an identity for the inverse of a sum of two square matrices, derived from the known formula for the 
determinant of such a sum.  

\end{abstract}
\maketitle
\section{Introduction}
\label{sect1}
Exactly solvable model problems play a central role in teaching as well as in ``intellectual body building" (John Negele). 
This is well documented in textbooks on basic subjects like classical mechanics, electrodynamics, thermodynamics, or quantum mechanics.
In advanced subjects like quantum field theory, it becomes increasingly 
difficult to identify such problems. Here one has to compromise, for example by resorting to a lower number of dimensions. In 1+1 
dimensions in particular, a number of quantum field theories are accessible by analytical means, either exactly or at least in certain limits.
A famous example is the Gross-Neveu (GN) model featuring self-interacting Dirac fermions with point interactions in 1+1 dimensions \cite{1}. Different
variants of this model are distinguished by their symmetries. In the case of continuous chiral symmetry, they are referred to as chiral GN models 
or 2d Nambu---Jona-Lasinio (NJL) models \cite{2}. The model of interest here belongs to this category and possesses the non-Abelian chiral symmetry group 
${\rm U}_L(N_f)\times{\rm U}_R(N_f)$. Its Lagrangian reads 
\begin{eqnarray}
{\cal L} & = & \sum_{k=1}^{N_c} \sum_{\alpha = 1}^{N_f} \bar{\psi}_{k,\alpha} i \partial \!\!\!/ \psi_{k,\alpha} +  \frac{g^2}{4}  \sum_{a=0}^{N_f^2-1} 
\left[ \left( \sum_{k=1}^{N_c}\sum_{\alpha = 1}^{N_f}  \bar{\psi}_{k,\alpha} (\lambda^a)_{\alpha \beta} \psi_{k,\beta}\right)^2 + \left(    \sum_{k=1}^{N_c} 
\sum_{\alpha=1}^{N_f} \bar{\psi}_{k,\alpha} i \gamma_5 (\lambda^a)_{\alpha \beta} \psi_{k,\beta} \right)^2 \right]
\nonumber \\
&  = &  \bar{\psi} i \partial \!\!\!/ \psi + \frac{g^2}{4}  \sum_{a=0}^{N_f^2-1} \left[ (\bar{\psi} \lambda^a \psi)^2 + (\bar{\psi} i \gamma_5 \lambda^a \psi)^2 \right].
\label{1.1}
\end{eqnarray}
The upper line shows explicitly the way in which ``color" indices ($k$) and ``flavor" indices $(\alpha,\beta)$ are contracted. Note that both 
of these refer to flavor in the present context, but color does not enter into the four-fermion interaction vertices. The lower line is the conventional shorthand
notation in this context where indices are suppressed whenever possible.
The $\lambda^a$ denote the $N_f^2-1$ SU($N_f$) generators in the fundamental representation with the usual normalization,
supplemented by $\lambda^0 = \sqrt{2/N_f}$ (proportional to a unit matrix) to account for U(1),
\begin{eqnarray}
{\rm Tr\, } \lambda^a \lambda^b & = &   2 \delta_{ab}, \quad a,b=0...N_f^2-1
\nonumber \\
\sum_{a=0}^{N_f^2-1}  \lambda^a_{\alpha \beta} \lambda^a_{\gamma\delta} & = &   \lambda^a_{\alpha \beta} \lambda^a_{\gamma \delta}  =   2 \delta_{\alpha \delta}\delta_{\beta \gamma}.
\label{1.2}
\end{eqnarray}
The prefactor $g^2/4$ in (\ref{1.1}) has been chosen such as to match the convention of the standard chiral GN model with ${\rm U}_L(1)\times{\rm U}_R(1)$  chiral symmetry,
\begin{equation}
{\cal L} = \bar{\psi} i \partial \!\!\!/ \psi + \frac{g^2}{2}  \left[ (\bar{\psi} \psi)^2 + (\bar{\psi} i \gamma_5  \psi)^2 \right] \quad (N_f=1).
\label{1.3}
\end{equation}
In 3+1 dimensions, Lagrangian (\ref{1.1}) is well known from the SU(3)-flavor version of the NJL model \cite{3}. There one usually adds a term which breaks U$_A$(1)
(the 't Hooft determinant), but we see no reason to do so in 1+1 dimensions. 
In order to validate a semiclassical approach, we shall only consider the 't Hooft limit $N_c \to \infty$, $N_c g^2$
= constant, in the present work. We do not include a bare fermion mass term, as this would prevent us from treating the problem analytically.
The basic semiclassical tools for a fermionic theory are the Hartree-Fock (HF) approach for static problems and the time dependent Hartree-Fock (TDHF) approach
for dynamical problems.

Studies of GN type field theories fall roughly into two categories: Thermodynamics and phase diagrams, or solitonic bound states and their interactions.
The issue which has perhaps received most attention in nuclear and particle physics are the phase diagrams as a function of temperature and chemical potentials.
In condensed matter physics on the other hand where closely related models arise in the context of quasi-one dimensional systems (polymers, superconductors, trapped ions), 
the focus has typically been on soliton spectra and dynamics. In this context, soliton refers to the behavior of the mean field, related to the fermionic single particle
wave functions by self-consistency. This type of bound state is a toy model for composite, relativistic objects, mimicking hadrons in real life.

In strong interaction physics and 
quantum chromodynamics (QCD), the method of choice has become the lattice Monte Carlo calculation in Euclidean space.
Unfortunately, neither of the issues just mentioned can be fully handled by this method. In the case of dense matter, the sign problem is still
a serious obstacle against working at finite chemical potential. Time dependent problems like hadron scattering can only indirectly be dealt with,
for instance by calculating scattering lengths with L\"uscher's method \cite{4}. 
These limitations make it desirable to gain some experience with interacting relativistic bound states of massless fermions in a reliable
way, complementary to numerical studies of lattice QCD. GN type models offer exact, analytical solutions for both phase diagrams and bound state dynamics,
if only in 1+1 dimensions.

In the present work, we reconsider the problem of soliton dynamics in the multiflavor chiral GN model (\ref{1.1}). To put our study into perspective, let us briefly recall
the state of the art of solving time dependent problems in GN models.

Static solitons have been found early on \cite{5} and have thoroughly been studied since then \cite{6}. The subject of soliton dynamics 
in the original GN model with discrete chiral symmetry also starts with Ref.~\cite{5} where the first time dependent mean field solution was found, the breather. This is a collectively  excited soliton,
vibrating in its rest frame. The authors guess its form by analogy with the sine-Gordon breather. They point out that it should
be related to kink-antikink scattering by analytic continuation. This idea was taken up again in Ref.~\cite{7} where kink-antikink scattering was solved  
in detail. The generalization to any number of colliding kinks followed soon afterwards \cite{8}.
The twisted kink in the one-flavor chiral GN model was discovered by Shei using inverse scattering theory \cite{9}. It is a ``chord soliton" in the sense that its mean field
traces out a straight line between two different vacua on the chiral circle. 
Bound states of an arbitrary number of twisted kinks were first found in Ref.~\cite{10} and generalized to time dependent scattering and breather phenomena in Ref.~\cite{11}. 
Compact analytical formulas are given for any number and complexity of solitons or breathers. More recently, these works have been further extended to the multiflavor case, 
first by using methods akin to inverse scattering theory in condensed matter physics \cite{12}. It turns out that the formalism of Ref.~\cite{13} from the particle physics side 
can also be generalized rather easily to $N_f$ flavors,
see Ref.~\cite{14} for the case $N_f=2$. As a matter of fact, the restriction to $N_f=2$ is unnecessary, as the formalism is practically independent of $N_f$. The fact that the $N_f$ 
flavor model can still be solved exactly is non-trivial. The static case was explored in more depth in Ref.~\cite{15}, a paper which has some overlap with the present one.

Summarizing, we already have at our disposal all the tools needed to compute bound and scattering states of twisted solitons or breathers. Exact expressions are available
for both mean field and spinors in closed analytical form. Self-consistency has been established quite generally. If the only goal was to present figures or animations
of specific collision events, this would be sufficient. However, in view of the pedagogical thrust of such studies, one would like to better understand what is going on, in 
particular concerning the role of flavor degrees of freedom. Ref.~\cite{15} has already provided us with useful additional insights for the static case, for instance for widely spaced solitons
and the vacua in between. Here we propose to extend this study to the most general time dependent solution, trying to disentangle flavor from the other degrees of freedom 
as much as possible. 

This paper is organized as follows. In Sect.~\ref{sect2}, we collect some basic facts about the chiral GN model with $N_f$ flavors. Sect.~\ref{sect3} briefly reviews the
analytical mean field solution for any number of solitons and/or breathers. We then proceed from there and transform the result for mean field and spinors into a
more illuminating form. To this end, we first had to derive an exact expression for the inverse of a sum of two matrices which may also be of interest for 
other purposes, see Sect.~\ref{sect4}. In Sect.~\ref{sect5} we apply it to the case where there are no breathers but only solitons (twisted kinks or bound states thereof).
The general case including breathers is the subject of Sect.~\ref{sect6}. This is followed by illustrative examples covering the single twisted kink, Sect.~\ref{sect7},
scattering or bound states of two kinks, Sect.~\ref{sect8}, and the breather, Sect.~\ref{sect9}. Finally, Sect.~\ref{sect10} contains a short summary and conclusions. The proof
of a mathematical identity is relegated to the appendix. 

\section{Basic facts about the model}
\label{sect2}
Lagrangian (\ref{1.1}) has an obvious U($N_c$) symmetry and a somewhat less obvious ${\rm U}_L(N_f) \times {\rm U}_R(N_f)$ chiral symmetry.
The latter becomes manifest once we decompose the spinors into left- and right-handed chiralities,
\begin{equation}
\psi = \frac{1+\gamma_5}{2} \psi_R + \frac{1-\gamma_5}{2} \psi_L.
\label{2.1}
\end{equation}
Using the representation
\begin{equation}
\gamma^0 = \sigma_1,\quad \gamma^1 = i \sigma_2, \quad \gamma_5 = \gamma^0 \gamma^1 = - \sigma_3
\label{2.2}
\end{equation}
of the Dirac matrices together with light cone coordinates
\begin{equation}
z=x-t, \quad \bar{z}=x+t, \quad \partial_0 = \bar{\partial}-\partial, \quad \partial_1 = \bar{\partial}+ \partial,
\label{2.3}
\end{equation}
we find
\begin{equation}
{\cal L} = 2i \psi_R^{\dagger} \bar{\partial} \psi_R - 2i \psi_L^{\dagger} \partial \psi_L + 2 g^2  (\psi_{L,\alpha}^{\dagger} \psi_{R,\beta})(\psi_{R,\beta}^{\dagger} \psi_{L,\alpha}).
\label{2.4}
\end{equation}
Again, color indices within bilinears are contracted to singlets. As U($N_f$) acts only on the flavor indices, the interaction term is manifestly ${\rm U}_L(N_f) \times {\rm U}_R(N_f)$ chirally invariant,
as is the free Lagrangian for massless Dirac fermions.
The Euler-Lagrange equations
\begin{eqnarray}
2 i \partial \psi_{L,\alpha} & = & 2 g^2  (\psi_{L,\alpha} \psi_{R,\beta}^{\dagger} ) \psi_{R,\beta},
\nonumber \\
2 i \bar{\partial} \psi_{R,\alpha} & = & - 2 g^2  (\psi_{R,\alpha} \psi_{L,\beta}^{\dagger} ) \psi_{L,\beta},
\label{2.5}
\end{eqnarray}
lead directly to the TDHF equations in the large $N_c$ limit,
\begin{eqnarray}
2 i \partial \psi_L & = & - \Delta^{\dagger} \psi_R,
\nonumber \\
2 i \bar{\partial} \psi_R & = & \Delta \psi_L.
\label{2.6}
\end{eqnarray}
Here, the mean field $\Delta$ is a color singlet, but a $N_f \times N_f$ matrix in flavor space. The self-consistency condition reads
\begin{equation}
\Delta_{\alpha \beta} = -2 N_c g^2 \langle \psi_{R,\alpha} \psi_{L,\beta}^{\dagger} \rangle
\label{2.7}
\end{equation}
where we have replaced bilinears by expectation values and pulled out the factor $N_c$ by performing trivial color sums. Due to the dyadic structure of $\Delta$,
it is now evident that both the TDHF equation and self-consistency are preserved under global chiral ${\rm U}_L(N_f)\times{\rm U}_R(N_f)$ transformations,
\begin{equation}
\psi_L \to U_L \psi_L, \quad \psi_R \to U_R \psi_R, \quad \Delta \to U_R \Delta U_L^{\dagger}.
\label{2.8}
\end{equation}
This observation can be used to simplify the vacuum problem. Due to spontaneous symmetry breakdown, the (homogeneous) vacuum is characterized by a constant matrix $\Delta$.
Using the freedom of performing global chiral transformations, we map it onto the unit matrix
times a scale factor,
\begin{equation}
\Delta_{\alpha \beta} = m \delta_{\alpha \beta} .
\label{2.9}
\end{equation}
The constant $m$ plays the role of dynamical fermion mass.
The HF vacuum energy density can be evaluated as the sum over single particle energies plus a double counting correction,
\begin{eqnarray}
{\cal E}_{\rm vac} & = & {\cal E}_{\rm sp}(m) + {\cal E}_{\rm dc}(m),
\nonumber \\
{\cal E}_{\rm sp} (m) & = & - N_c N_f \int_{-\Lambda/2}^{\Lambda/2} \frac{dk}{2\pi} \sqrt{k^2+m^2}
\nonumber \\
& = & - N_c N_f \left[ \frac{\Lambda^2}{8\pi} - \frac{m^2}{4\pi} \left( \ln \frac{m^2}{\Lambda^2} -1 \right) \right],
\nonumber \\
{\cal E}_{\rm dc}(m) & = &  N_f \frac{m^2}{2 g^2}.
\label{2.10}
\end{eqnarray}
The gap equation follows by minimizing ${\cal E}_{\rm vac}$ with respect to $m$,
\begin{equation}
0 = 1 + \frac{N_c g^2}{2\pi} \ln \frac{m^2}{\Lambda^2},
\label{2.11}
\end{equation}
and is indistinguishable from that of the one-flavor model.
The renormalized vacuum energy density
\begin{equation}
{\cal E}_{\rm vac} = - N_c N_f \frac{m^2}{4\pi}
\label{2.12}
\end{equation}
is just $N_f$ times the known energy density of the one-flavor model.
The fermion mass $m$ which arises from dimensional transmutation can be set equal to 1 by choice of units. This is what we shall do in the present work.
The vacuum manifold then coincides with the group U($N_f$), as can be seen by applying all global chiral transformations to $\Delta=1$. For one flavor, it reduces
to U(1), i.e., the familiar chiral circle.

\section{Reminder of the multisoliton solution}
\label{sect3}
In the large $N_c$ limit, it is possible to solve bound state and scattering problems in model (\ref{1.1}) explicitly. This holds for any number of solitons 
and/or breathers as well as for any $N_f$. Originally, a general solution has been found for the one-flavor model. In Ref.~\cite{12} (in condensed matter physics) and \cite{14} (in particle physics), 
this solution was subsequently generalized to several flavors. In Ref.~\cite{14} in particular, the $N_f=2$ case has been treated in some detail. Since the generalization to 
arbitrary $N_f$ is trivial, we infer the solution from this work and present only the necessary definitions and results, referring to \cite{14} for detailed proofs.
The upcoming sections of the present work may then be regarded as an elaboration on this general solution. 

The starting point for attacking $N$ soliton problems is a $N$-dimensional vector $e$ 
with components \cite{13}
\begin{equation}
e_i = e^{i (\zeta_i^* \bar{z} -z/\zeta_i^*)/2}.
\label{3.1}
\end{equation}
The $\zeta_i$ are complex numbers (Im $\zeta_i >0$) characterizing the pole positions of the TDHF continuum
wave functions in the complex $\zeta$ plane. Here, $\zeta$ is the spectral parameter
related to light cone momentum and energy (``uniformizing parameter" in condensed matter language), 
\begin{equation}
k=\frac{1}{2} \left( \zeta- \frac{1}{\zeta}\right), \quad E= - \frac{1}{2} \left( \zeta + \frac{1}{\zeta} \right).
\label{3.2}
\end{equation}
Note that
\begin{equation}
k_{\mu}x^{\mu} = - \frac{1}{2} \left( \zeta \bar{z} - \frac{z}{\zeta} \right),
\label{3.3}
\end{equation}
so that $e_i$ is recognized as a plane wave evaluated at a complex spectral parameter corresponding to a bound state pole.
The crucial step when going from one to $N_f$ flavors consists in dressing each $e_i$ with a flavor vector $\vec{p}_i$ 
\begin{equation}
e_i \to e_i \vec{p}_i , \quad e_{i,\alpha} = e_i p_{i,\alpha}  \quad ({\rm no\,\,} i{\rm -sum})
\label{3.4}
\end{equation}
Here, $\vec{p}_i$ is a $N_f$-component, constant, complex vector with components $p_{i,\alpha}$. 
Its precise meaning will be clarified later on. Since a common real factor multiplying $\vec{p}_i$ can always be absorbed in the soliton positions,
we can assume that these vectors are normalized ($\vec{p}_i^{\,\dagger} \vec{p}_i=1$) without loss of generality.
However they are in general neither orthogonal nor even linearly independent. This is obvious since the number
of solitons may exceed the number of flavors.
We continue using Greek indices for flavor and suppress the indices $i=1...N$ referring to the bound state poles whenever possible.
Then we have to modify the results of Refs.~\cite{13} as follows:
Continuum TDHF spinors are now 2$N_f$-component objects 
\begin{equation}
\psi_{\zeta,\alpha} = \frac{1}{\sqrt{1+\zeta^2}} \left( \begin{array}{c} \zeta \chi_{1,\alpha} \\ - \chi_{2,\alpha} \end{array} \right) e^{i(\zeta \bar{z}-z/\zeta)/2}.
\label{3.5}
\end{equation}
The following ansatz for the $\chi_{i,\alpha}$ is motivated by the assumed pole structure of the continuum spinors ($N$ poles, corresponding to $N$ bound states),
\begin{eqnarray}
\chi_{1,\alpha} & = & \left( \delta_{\alpha \beta} + i \sum_{i=1}^N \frac{1}{\zeta-\zeta_i}  \varphi_{1,i,\alpha} e_{i,\beta}^* \right) q_{\beta},
\nonumber \\
\chi_{2,\alpha} & = & \left( \delta_{\alpha \beta} - i \sum_{i=1}^N \frac{\zeta}{\zeta-\zeta_i}  \varphi_{2,i,\alpha} e_{i,\beta}^* \right) q_{\beta}.
\label{3.6}
\end{eqnarray}
The $q_{\beta}$ are the amplitudes of the flavor components of the incoming plane wave
\begin{equation}
\psi_{\zeta,\alpha}|_{\rm in} = \frac{1}{\sqrt{1+\zeta^2}} \left( \begin{array}{c} \zeta \\ -1 \end{array} \right) e^{i(\zeta \bar{z}- z/\zeta)/2} q_{\alpha}.
\label{3.7}
\end{equation}
The vacuum at $x\to - \infty$ will always be chosen as $\Delta_{\rm vac}=1$. 
When summing over all continuum states, the $\vec{q}$ should be chosen in all flavor directions (one component 1, all the others 0) to account
for incoming waves in the different flavor channels. The $\varphi_{1,i,\alpha},\varphi_{2,i,\alpha}$ introduced in Eq.~(\ref{3.6}) are closely related to the components of bound state spinors.
They can be evaluated by linear algebra as follows: Define a hermitean $N \times N$ matrix $B$, 
\begin{equation}
B_{ij}= i \frac{e_{i,\beta} e_{j,\beta}^*}{\zeta_j-\zeta_i^*} = i \frac{e_i e_j^*}{\zeta_j-\zeta_i^*} \vec{p}_j^{\,\dagger}\vec{p}_i .
\label{3.8}
\end{equation}
The $\varphi_{1,i,\alpha},\varphi_{2,i,\alpha}$ then satisfy the following system of linear, algebraic equations
\begin{eqnarray}
(\omega + B) \varphi_{1,\alpha} & = & e_{\alpha},
\nonumber \\
(\omega + B) \varphi_{2,\alpha} & = & - f_{\alpha},
\label{3.9}
\end{eqnarray}
where $f_{i,\alpha}=e_{i,\alpha}/\zeta_i^*$. 
Like in the one flavor case, a constant, hermitean $N \times N$ matrix $\omega$ encoding further information
about the soliton configuration (geometry,  initial conditions, breather frequency and amplitude) has been introduced.
The dimension of the linear system (\ref{3.9}) does not increase with the number of flavors, but depends only on the total number of bound state poles.
What is new as compared to the one-flavor case is the factor $\vec{p}_j^{\,\dagger}\vec{p}_i$ in $B_{ij}$ and the fact
that one gets a pair of linear equations for each flavor component $\alpha$. 
The most important result for the following is the expression for the mean field, now a $N_f \times N_f$ matrix 
\begin{equation}
\Delta_{\alpha  \beta} =  \delta_{\alpha \beta}+i e_{\beta}^{\dagger} \frac{1}{\omega+B} f_{\alpha}.
\label{3.10}
\end{equation}
Orthonormal bound states can be constructed as in the one-flavor case by linear combinations of the $\varphi_i$,
\begin{equation}
\hat{\varphi}_i = \sum_j\ C_{ij} \varphi_j, \quad \int dx \hat{\varphi}_{i, \alpha}^{\dagger}\hat{\varphi}_{j,\alpha} = \delta_{ij}.
\label{3.11}
\end{equation}
The resulting condition coincides with the one in the one-flavor case,
\begin{equation}
2 C \omega^{-1} C^{\dagger} = 1.
\label{3.12}
\end{equation}
A central ingredient of the TDHF calculation is the self-consistency condition. We introduce two diagonal $N\times N$ matrices
\begin{eqnarray}
M_{ij} & = &  -i \delta_{ij} \ln(-\zeta_i^*),
\nonumber  \\
N_{ij} & = &  4 \pi \delta_{ij} \nu_i,
\label{3.13}
\end{eqnarray}
where $\nu_i$ is the occupation fraction of bound state $i$. The self-consistency condition then assumes the form
\begin{equation}
\omega M^{\dagger} + M \omega = C^{\dagger} N C
\label{3.14}
\end{equation}
independently of the number of flavors.

The following observations carry over from the one-flavor to the multiflavor models.
Owing to chiral symmetry, model (\ref{1.1}) gives rise to $N_f^2$ conserved vector and axial vector Noether currents 
\begin{eqnarray}
\partial ^{\mu} j^a_{\mu} & = & \partial^{\mu} (\bar{\psi} \gamma_{\mu} \lambda^a \psi) = 0
\nonumber \\
\partial ^{\mu} j^a_{5,\mu} & = & \partial^{\mu} (\bar{\psi} \gamma_{\mu}\gamma_5  \lambda^a \psi) = 0
\label{3.15}
\end{eqnarray}
In 1+1 dimensions, vector and axial vector currents are not independent, but satisfy
\begin{equation}
j_5^{a,0}  =  j^{a,1}, \quad j_5^{a,1} = j^{a,0},
\label{3.16}
\end{equation}
Adding and subtracting the conservation laws (\ref{3.15}) and introducing light cone coordinates (\ref{2.3}),
one finds
\begin{equation}
\bar{\partial} (\psi_R^{\dagger}\lambda^a \psi_R)=0, \quad \partial (\psi_L^{\dagger}\lambda^a \psi_L) = 0.
\label{3.17}
\end{equation}
If we take the expectation value of these equations in an arbitrary state, we conclude that the right-handed density $\rho_R^a  = \langle \psi_R^{\dagger}\lambda^a \psi_R \rangle$
depends only on $z=x-t$, the left-handed density $\rho_L^a  = \langle \psi_L^{\dagger}\lambda^a \psi_L \rangle$ only on $\bar{z}=x+t$, i.e., they can only move with the speed of light
to the right or to the left (or be constant). In a localized, massive state like a solitonic bound state or breather, these densities must therefore vanish identically. 
Hence we anticipate that all densities and current densities must vanish inside an arbitrary
soliton or multisoliton state, at least in the strict thermodynamic and chiral limit. This should hold for left- and right-handed fermions separately, or, equivalently, for charge and current densities.
This can indeed be verified by a detailed computation (see Ref.~\cite{14}) and holds for all flavor currents including the fermion current (the $a=0$ component).

The formulas given in the present section are sufficient for computing the space-time evolution of any multisoliton event. However, it turns out that one can convert the result into a more transparent and useful form, notably in the 
absence of breathers. Ref.~\cite{15} has already dealt with the static case in a similar spirit. The goal of the following sections is to simplify the time dependent case as well.
Thus we shall start from Eq.~(\ref{3.10}) for $\Delta$ and transform it to a more instructive expression. The key problem here is how to invert the matrix $(\omega+B)$. Once this has been achieved,
the formulas for the spinors can be simplified as well.

\section{Inverting a sum of two matrices}
\label{sect4}

Let us go back to Eq.~(\ref{3.10}) for the mean field,
\begin{equation}
\Delta_{\alpha \beta} = \delta_{\alpha \beta} + i e_{\beta}^{\dagger} \frac{1}{\omega+B} f_{\alpha},
\label{4.1}
\end{equation}
with
\begin{eqnarray}
B_{ij}   & = &   i \frac{e_i e_{j}^*}{\zeta_j-\zeta_i^*} \sigma_{ji}, \quad \sigma_{ji}= \vec{p}_j^{\,\dagger} \vec{p}_i,
\nonumber \\
e_{i, \alpha} & = & e_i p_{i,\alpha}, \quad f_{i,\alpha} = (\zeta_i^*)^{-1} e_{i,\alpha}.
\label{4.2}
\end{eqnarray}
The constant matrix $\omega$ is diagonal for the scattering case (solitons, bound states) and off diagonal for problems involving breathers. 
Although expression (\ref{4.1}) is exact, it is not yet very transparent. Space-time-, flavor- and parameter dependences are inextricably entangled.
We therefore pull out the $(x,t)$ dependent factors from $B$ by introducing a diagonal $N \times N$ matrix $E$ as follows
\begin{eqnarray}
B & = &  E S E^{\dagger},
\nonumber \\
E_{ij} & = & \delta_{ij} e_i,
\nonumber \\
S_{ij} & = & \frac{i}{\zeta_j-\zeta_i^*} \sigma_{ji}, \quad S^{\dagger}=S.
\label{4.3}
\end{eqnarray}
Thus
\begin{equation}
\omega+B   =    E ( v + S ) E^{\dagger}, \quad v  =  E^{-1}\omega (E^{\dagger})^{-1} .
\label{4.4}
\end{equation}
Inserting this expression into (\ref{4.1}) yields
\begin{equation}
\Delta_{\alpha \beta} = \delta_{\alpha \beta} + i e_{\beta}^{\dagger}(E^{\dagger})^{-1}  \frac{1}{v+S} E^{-1} f_{\alpha}.
\label{4.5}
\end{equation}
In the denominator, the ($x,t$) dependence has now been shifted to the matrix $v$. 
The diagonal matrices $E^{-1}, (E^{\dagger})^{-1}$ acting on the vectors $f_{\alpha}, e^{\dagger}_{\beta}$ cancel the ($x,t$) dependence of these
vertex functions. Inserting $e_{\beta}^{\dagger}$ and $ f_{\alpha}$, $\Delta$ is expressed in flavor space
via a sum over dyadics $\vec{p}_j \vec{p}_i^{\,\dagger}$ ,
\begin{equation}
\Delta = 1  + i \sum_{i,j}  \left( \frac{1}{v+S}\right)_{ij} \frac{1}{\zeta_j^*} \vec{p}_{j}\vec{p}_i^{\,\dagger}.
\label{4.6}
\end{equation}
The coefficients of $\vec{p}_j \vec{p}_i^{\,\dagger}$ require inverting the sum of a hermitean, space-time dependent matrix $v$
and a hermitean, constant matrix $S$. The explicit flavor dependence is through the dyadics $\vec{p}_j \vec{p}_i^{\, \dagger}$. In addition,
there is an implicit flavor dependence through $S$ which still depends on the flavor scalars $\sigma_{ij}=\vec{p}_i^{\,\dagger} \vec{p}_j$. 

Along the same lines, we rewrite the expressions for the spinors in terms of the inverse of the matrix ($v+S$). Using once again arrows for vectors in flavor space, we find
\begin{eqnarray}
\vec{\varphi}_{1,i} & = & \sum_j \frac{1}{e_i^*} \left( \frac{1}{v+S} \right)_{ij} \vec{p}_j,
\nonumber \\
\vec{\varphi}_{2,i} & = & - \sum_j \frac{1}{e_i^*} \left( \frac{1}{v+S} \right)_{ij} \frac{1}{\zeta_j^*}  \vec{p}_j,
\nonumber \\
\vec{\chi}_{1,\vec{q}} & = &  \left( 1+ i \sum_{i,j} \frac{1}{\zeta-\zeta_i} \left(  \frac{1}{v+S} \right)_{ij} \vec{p}_j \vec{p}_i^{\,\dagger} \right) \cdot \vec{q},
\nonumber \\
\vec{\chi}_{2,\vec{q}} & = &  \left( 1+ i \sum_{i,j}  \frac{\zeta}{\zeta-\zeta_i} \left( \frac{1}{v+S} \right)_{ij}\frac{1}{\zeta_j^*}  \vec{p}_j \vec{p}_i^{\,\dagger} \right) \cdot \vec{q}.
\label{4.7}
\end{eqnarray}
The full continuum spinors become
\begin{equation}
\vec{\psi}_{\zeta,\vec{q}}  =  \frac{1}{\sqrt{1+\zeta^2}} \left( \begin{array}{c} \zeta \vec{\chi}_{1,\vec{q}} \\ - \vec{\chi}_{2,\vec{q}} \end{array} \right) e^{ i(\zeta \bar{z}-z/\zeta) /2} .
\label{4.8}
\end{equation}
Common to all expressions (\ref{4.6},\ref{4.7}) is the appearance of the inverse matrix $(v+S)^{-1}$. Since all the space-time dependence is now in $v$, this raises 
the question about a useful expression where one can keep track of $v$ and $S$ separately also in $1/(v+S)$.  Since we could not find an appropriate formula
in the literature, we first derive a general expression for the inverse of a sum of two square matrices. 

We start from a known expression for the determinant of a sum of matrices \cite{16}. It reads as follows:
\begin{equation}
\det (A+B) = \det A + \det B + \sum_{r=1}^{N-1}  \sum_{\alpha,\beta} (-1)^{s(\alpha)+s(\beta)} \det  A[\alpha | \beta ] \det  B(\alpha | \beta ) .
\label{4.9}
\end{equation}
$A$ and $B$ are $N$-square matrices. The outer sum is over integers $r$ from 1 to $N-1$. For a particular $r$, the inner sum is over 
all strictly increasing integer sequences $\alpha$ and $\beta$ of length $r$ chosen from $1,...,N-1$. $A[\alpha | \beta]$ (square brackets)
is the $r$-square submatrix of $A$ lying in the rows $\alpha$ and columns $\beta$. $B(\alpha |\beta)$ (round brackets) is the ($N-r$)-square
submatrix of $B$ lying in rows complementary to $\alpha$ and columns complementary to $\beta$. $s(\alpha)$ is the sum of all integers in $\alpha$. 

We are interested in the inverse matrix $(A+B)^{-1}$, i.e.
\begin{equation}
\frac{1}{A+B} = \frac{{\rm adj}(A+B)}{\det (A+B)}.
\label{4.10}
\end{equation}
Here, adj denotes the classical adjoint (or adjugate) matrix, the transpose of the cofactor matrix. 
The denominator is taken care of by Eq.~(\ref{4.9}). For the numerator, we find a similar equation
\begin{equation}
{\rm adj}(A+B) = {\rm adj\,} B + \sum_{r=1}^{N-1}   \sum_{\alpha,\beta} (-1)^{s(\alpha)+s(\beta)} \det  A[\alpha | \beta ]  \widetilde{{\rm \,adj} B(\alpha | \beta ) }
\label{4.11}
\end{equation}
where one of the det-factors in the sum has been replaced by a new symbol. 
The tilde above adj $B(\alpha|\beta)$ has the following meaning: Evaluate the ($N-r$)-square matrix $B(\alpha|\beta)$ and take its adjoint, then ``inflate" the result to a $N$-square
matrix by filling rows $\beta$ and columns $\alpha$ with zero's [note the interchange of rows and columns
as compared to the definition of $B(\alpha|\beta)$]. Incidentally, the $r=N-1$ term in (\ref{4.11}) is equal to adj $A$.

Actually, Eq.~(\ref{4.11}) is a consequence of (\ref{4.9}). To show this, 
consider the ($i,j$)-matrix element of Eq.~(\ref{4.11}). Left hand side,
\begin{equation}
({\rm l.h.s.})_{ij} = \left[ {\rm adj}(A+B) \right]_{ij} = (-1)^{i+j} \det \left[ A(j|i) + B(j|i) \right].
\label{4.12}
\end{equation}
The determinant of the sum $A(j|i)+B(j|i)$ in turn can again be evaluated with the help of Eq.~(\ref{4.9}) for $N \to N-1$. In order to apply the formula literally, the row and column
indices should run from 1 to $N-1$. Now the indices run from 1 to $N$ with $j$ missing in the row indices and $i$ missing in the column indices. This does not
affect the determinants of submatrices in (\ref{4.9}), but it does affect the phase factor. The correct result is
\begin{eqnarray}
\det \left[ A(j|i) + B(j|i) \right] & = & \det  A(j|i)   + \det  B(j|i) 
\label{4.13} \\
&+ &  \sum_{r=1}^{N-2} \sum_{\bar{\alpha}_j,\bar{\beta}_i} (-1)^{s(\bar{\alpha}_j)+s(\bar{\beta}_i)+n(\bar{\alpha}_j) +n(\bar{\beta}_i)} \det  A[\bar{\alpha}_j|\bar{\beta}_i]  \det B(j,\bar{\alpha}_j|i,\bar{\beta}_i) 
\nonumber
\end{eqnarray}
The $\bar{\alpha}_j,\bar{\beta}_i$ are defined like the $\alpha,\beta$ above except that $\bar{\alpha}_j$ does not contain the index $j$, $\bar{\beta}_i$ does not contain the index $i$. Their maximal
length is thereby reduced to $N-2$. Furthermore, $n(\bar{\alpha}_j)$ is the number of elements of the sequence $\bar{\alpha}_j$ that are $> j$, $n(\bar{\beta}_i)$ the number of elements of $\bar{\beta}_i$ that are $> i$.
These modifications of the phase factor are necessary because the labeling of rows and columns in $A(j|i),B(j|i)$ is not the standard one.

On the right hand side of (\ref{4.11}), we split the index set $\alpha$ into sets $\alpha_j$ containing $j$ and sets $\bar{\alpha}_j$ not containing $j$. Similarly
for $\beta$, $\beta_i$ (containing $i$) and $\bar{\beta}_i$ (not containing $i$). The summation over $\alpha,\beta$ then gives rise to 4 terms
\begin{equation}
\sum_{\alpha,\beta} \to \sum_{\alpha_j,\beta_i} + \sum_{\alpha_j,\bar{\beta}_i} + \sum_{\bar{\alpha}_j,\beta_i} + \sum_{\bar{\alpha}_j,\bar{\beta}_i} .
\label{4.14}
\end{equation}
If we take the ($i,j$)-matrix element of $\widetilde{{\rm adj}(B(\alpha|\beta))}$ in Eq.~(\ref{4.11}), only the last term in (\ref{4.14}) contributes since the rows filled with zero's are given by $\beta$, the columns 
filled with zero's by $\alpha$. To get a non-zero row index $i$, $\beta$ should not contain $i$, hence $\bar{\beta}_i$ is needed. To get a non-zero column index $j$, $\alpha$ should not contain $j$, 
hence $\bar{\alpha}_j$ is needed.
The ($i,j$)-matrix element of $\widetilde{{\rm adj} B(\bar{\alpha}_j|\bar{\beta}_i)}$ in (\ref{4.11}) is given by
\begin{equation}
\left[  \widetilde{{\rm adj}B(\bar{\alpha}_j|\bar{\beta}_i}   \right]_{ij}     = (-1)^{i+j+n(\bar{\alpha}_j) +n(\bar{\beta}_i)} \det  B(j,\bar{\alpha}_j|i,\bar{\beta}_i)  .
\label{4.15}
\end{equation}
The extra phase factor is again due to the non-standard labeling of rows and columns in the submatrices. This comes about as follows. When filling the rows labeled by $\bar{\beta}_i$ with zero's, the row index 
increases by the number of elements of $\bar{\beta}_i$ that are $<i$, i.e., $r-n(\bar{\beta}_i)$. When filling the columns labeled by $\bar{\alpha}_j$ with zero's, the column index 
increases by the number of elements of $\bar{\alpha}_j$ that are $<j$, i.e., $r-n(\bar{\alpha}_j)$. This is the reason behind the phase factor in (\ref{4.14}). 
In addition, we pick up a term from ${\rm adj\,} B$ in (\ref{4.11}). 
Putting everything together, the ($i,j$)-matrix element of the right hand side of (\ref{4.11}) becomes identical to the left hand side as given in Eqs.~(\ref{4.12},\ref{4.13}).

Summarizing, we write down the full expression for the inverse of a sum of two matrices, expressing the right hand side by submatrices. To clarify
the idea behind this expression, we multiply $A$ by a formal parameter $\epsilon$. The terms of the sum then go like $\epsilon^r$. Thus one can think of
the formula as representing an exact expression for $1/(\epsilon A+  B)$ as a rational function in $\epsilon$, to be contrasted to the power series expansion
familiar from perturbation theory,
\begin{equation}
\frac{1}{ \epsilon A+ B} = \frac{ {\rm adj\,} B + \sum_{r=1}^{N-2} \epsilon^r \sum_{\alpha,\beta} (-1)^{s(\alpha)+s(\beta)} \det A[\alpha|\beta]\, \widetilde{{\rm adj\,} B(\alpha|\beta)} +\epsilon^{N-1} {\rm adj\,} A }
{\det B  + \sum_{r=1}^{N-1} \epsilon^r \sum_{\alpha,\beta} (-1)^{s(\alpha)+s(\beta)} \det A[\alpha|\beta] \det B(\alpha|\beta)  +\epsilon^N \det A }.
\label{4.16}
\end{equation}
The numerator (denominator) is a polynomial in $\epsilon$ of degree $N-1$ ($N$). The coefficients are given explicitly in terms of determinants and adjoints of submatrices of $A$ and $B$. 

An important special case for our purpose is the case where $A$ is diagonal. This covers all multisoliton interactions without breathers. Let us simplify the notation and formulas for this particular case.
Assume that
\begin{equation}
A_{ij}=\delta_{ij} A_i.
\label{4.17}
\end{equation}
Then in both Eqs.~(\ref{4.9}) and (\ref{4.11}) only $\alpha=\beta$ appears, so that the phase factor drops out. Using the simplified notation
\begin{equation}
A[\alpha|\alpha] = A[\alpha], \quad B(\alpha|\alpha) = B(\alpha)  ,
\label{4.18}
\end{equation}
we get
\begin{eqnarray}
\det(A+B) & = & \det A + \det B + \sum_{\alpha \in K_{N-1}}\det A[\alpha] \det B(\alpha),
\nonumber \\
{\rm adj} (A+B) & = & {\rm adj\,} B + \sum_{\alpha \in K_{N-1}} \det  A[\alpha] \, \widetilde{{\rm adj\,} B(\alpha)} ,
\label{4.19}
\end{eqnarray}
where $K_n$ denotes the set of all sequences of the type $\alpha$ as defined above with length $r=1,..,n$.
Moreover, the determinants of $A$ and its submatrices simplify to
\begin{equation}
\det A = \prod_{i=1}^N A_i, \quad 
\det A[\alpha] = \prod_{i \in \alpha} A_i.
\label{4.20}
\end{equation}

\section{Multisoliton dynamics without breathers}
\label{sect5}

If we disregard breathers for the moment, the matrix $\omega$ is diagonal,
\begin{equation}
\omega_{ij} = \delta_{ij} \omega_i.
\label{5.1}
\end{equation}
According to (\ref{4.4}), $v$ is also diagonal,
\begin{equation}
v_{ij} = \delta_{ij} v_i, \quad v_i = \frac{\omega_i}{|e_i|^2} = \frac{1}{V_i},
\label{5.2}
\end{equation}
where the $V_i$ are the basic profile functions of the solitons familiar from the one-flavor case \cite{13}. Their explicit form will be given below when we discuss examples of few soliton problems.
Recall that the mean field $\Delta$ in the single flavor case could be represented as a ratio of two multivariate polynomials in the 
$V_i$. Our goal is a corresponding expression for the multiflavor case. Since $\Delta$ is a flavor matrix, we expect the coefficients in the numerator to be flavor matrices as well.
In fact, Eq.~(\ref{4.19}) is exactly what is needed for this purpose. If we replace $v_i$ by $1/V_i$, the formula gives us directly the numerator and denominator of
$\Delta$ as polynomials in the $V_i$. For the $N$ soliton problem, matrix inversion needs to be done only for constant square matrices $S(\alpha)$ of dimension $N$ and lower.
The space-time dependence is contained in the monomials of $V_i$'s. By contrast, the original expression for $\Delta$, Eq.~(\ref{4.1}), requires matrix inversion for every $(x,t)$. Besides, 
as we shall see shortly, the structure of the result is more transparent and can be used to gain further analytical insights.

Let us introduce the $V_i$ at this stage. If we multiply numerator and denominator of $(v+S)^{-1}$ by $\det V= 1/\det v =  V_1...V_N$, we find that $\det V[\alpha]$ 
now multiplies the determinant or the adjoint of the matrix $S$ where the rows and columns in $\alpha$ are kept, rather than being removed
[$S[\alpha]$ in the notation of Eq.~(\ref{4.18})]. Thus
\begin{equation}
(v+S)^{-1} = \frac{\sum_{\alpha \in K_N} \det  V[\alpha] \, \widetilde{ {\rm adj\,} S[\alpha]} }
{1+   \sum_{\alpha \in K_N} \det  V[\alpha] \det S[\alpha] }
\label{5.3}
\end{equation}
The tilde above adj $S[\alpha]$ now instructs us to fill all rows and columns complementary to the set $\alpha$ with zero's. Unlike Eq.~(\ref{4.19}),
the sum over $\alpha$ includes the term of length $r=N$, with $V[1..N]=V$, $S[1..N]=S$. This is allowed here since there are no submatrices 
with round brackets of type $B(\alpha)$ which would be ill-defined for $r=N$. Also note that the term in the numerator with $\alpha$ of length 1 equals $V$.
After inserting (\ref{5.3}) into expression (\ref{4.6}) for $\Delta$, we arrive at the final result for the mean field,  
\begin{equation}
\Delta = \frac{1 +\sum_{\alpha \in K_N} \xi_{\alpha} [V]_{\alpha} U_{\alpha}}{1 + \sum_{\alpha \in K_N} \xi_{\alpha} [V]_{\alpha}}
\label{5.4}
\end{equation}
with the shorthand notation
\begin{eqnarray}
\xi_{\alpha} & = & \det S[\alpha] ,
\nonumber \\
\ [V]_{\alpha} & = &  \det  V[\alpha] = \prod_{i \in \alpha} V_i,
\nonumber \\
U_{\alpha} & = & 1 + i \sum_{i,j} \left( \widetilde{ S[\alpha]^{-1}}\right)_{ij} \frac{1}{\zeta_j^*} \vec{p}_j \vec{p}_i^{\,\dagger}.
\label{5.5}
\end{eqnarray}
We have replaced adj$\,S[\alpha]$ by $S[\alpha]^{-1}$ times $\det S[\alpha] = \xi_{\alpha}$ in $U_{\alpha}$.

The result (\ref{5.4},\ref{5.5})  is very simple indeed. Every single term in the numerator has the same structure, except that the indices are constrained to 
the sequence $\alpha$. One can check that this result reduces to the known expression in the one-flavor case.
To this end, treat $\vec{p}_i$ as a 1-component object with $p_i=1$ for all $i$. Since all $\sigma_{ij}=1$, the matrix $S[\alpha]$ becomes a Cauchy matrix 
for which the determinant and the inverse matrix are explicitly calculable \cite{17}. 
In the multiflavor case, the matrices $U_{\alpha}$ are unitary, as we will now check. For notational simplicity, consider $U:=U_{1...N}$,
the asymptotic vacuum at $x\to \infty$. The other $U_{\alpha}$'s can be handled similarly by merely restricting and relabeling the soliton indices.
We start from  
\begin{eqnarray}
U & = & 1 + i \sum_{n,j} \left(  S^{-1}\right)_{nj} \frac{1}{\zeta_j^*} \vec{p}_j \vec{p}_n^{\,\dagger},
\nonumber \\
U^{\dagger}&  = &  1-i \sum_{i,m} (S^{-1})_{im}\frac{1}{\zeta_i} \vec{p}_m \vec{p}_i^{\,\dagger},
\label{5.6}
\end{eqnarray}
where we have used the hermiticity of $S$. Computing $U U^{\dagger}$ yields a 1, two terms linear and one 
term quadratic in the dyadics. The quadratic term just cancels the sum of the linear terms. To show this, use
\begin{equation}
(\vec{p}_j \vec{p}_n^{\,\dagger} )(\vec{p}_m \vec{p}_{i}^{\,\dagger}) = \sigma_{n m} \vec{p}_j \vec{p}_i^{\,\dagger},
\label{5.7}
\end{equation}
express $\sigma_{nm}$ by $S_{nm}$, Eq.~(\ref{4.3}), and verify the identity
\begin{equation}
(S^{-1})_{nj} (S^{-1})_{im} \sigma_{nm} = -i (S^{-1})_{ij} ( \zeta_i-\zeta_j^*) .
\label{5.8}
\end{equation}
Along the same lines, we could insert our expression for $(v+S)^{-1}$ into the spinors, Eq.~(\ref{4.7}), in the compact form
\begin{equation}
\left( \frac{1}{v+S} \right)_{ij} = \frac{ \sum_{\alpha \in K_N} \xi_{\alpha} [V]_{\alpha} \left( \widetilde{ S[\alpha]^{-1}} \right)_{ij}}{1 + \sum_{\alpha \in K_N} \xi_{\alpha} [V]_{\alpha}}.
\label{5.9}
\end{equation}
Since the gain of insight is less obvious than in the case of $\Delta$, we shall not write down the resulting expressions here.

The $U_{\alpha}$'s have a simple physics interpretation. This becomes particularly clear if we assume that the solitons are all well separated, either by choice of bound state configuration (see \cite{15}) 
or at a certain time during a scattering process. Let us assume that the positions of the solitons are ordered such that $x_i \ll x_j \ll x_k ...$. In between two neighboring solitons, the mean field must reduce to 
that of a twisted vacuum, characterized by a locally constant matrix $\in {\rm U}(N_f)$. Proceeding from $x\to - \infty$ towards $x \to \infty$, the vacua are ordered as $1, U_i, U_{ij}, U_{ijk},... U_{12..N}$. The fact that all permutations
can occur is responsible for the proliferation of $U_{\alpha}$'s in $\Delta$, necessary to account for all possible orderings. In the one-flavor case, the $V_i$ are the same as here but the $U_{\alpha}$ go over into the familiar twist factors.
This follows from the remarkable identity for the Cauchy matrix $S$,
 \begin{equation}
U_{\alpha} \to  1 + i \sum_{i,j} \left( S[\alpha]^{-1}\right)_{ij} \frac{1}{\zeta_j^*} = \prod_{k\in \alpha} \frac{\zeta_k}{\zeta_k^*} \in {\rm U}(1) \quad (N_f=1).
\label{5.10}
\end{equation}
The prefactors $\xi_{\alpha}$ also depend on $N_f$, since the matrix $S$ contains the scalar products $\vec{p}_i^{\, \dagger} \vec{p}_j$. Thus we can think of the $U_{\alpha}$ either as twist matrices,
or as possible vacua far away from the solitons.
If two solitons are close together, the mean field in between is no longer related to a vacuum, but becomes ($x,t$)-dependent. However, expression (\ref{5.4}) remains valid.

The picture of widely separated solitons which emerges involves a sequence of solitons connecting the vacua
$1 \to U_1 \to U_{12} \to U_{123}...$ and all permutations. The intrinsic form of a single kink connects $1\to U_i$. By a chiral
transformation, we may identify intrinsic solitons with the sequence $1 \to U_1$, $1 \to U_{12} U_1^{\dagger}$, $1 \to U_{123} U_{12}^{\dagger}$ etc.
This corresponds to the decomposition of the vacua according to
\begin{eqnarray}
U_{12} & = & (U_{12}U_1^{\dagger}) U_1,
\nonumber \\
U_{123} & = & (U_{123}U_{12}^{\dagger}) (U_{12}U_1^{\dagger}) U_1,
\label{5.11}
\end{eqnarray}
etc. Each factor is an elementary twist matrix. It is of some interest to evaluate these elementary twist matrices, since they clearly show how a soliton is influenced by the flavor structure 
of the other twisted kinks. It is sufficient to compute $U U_k^{\dagger}$ for this purpose, where $U=U_{12..N}$ is the highest $U$
and $U_k$ differs from $U$ by the missing $k$-th row and column. All other products in (\ref{5.11})  can be obtained by restricting and renaming the indices. Here we only give the result,
referring to the appendix for the derivation. Let
\begin{eqnarray}
U & = & 1 + i \sum_{i,j} (S^{-1})_{ij} \frac{1}{\zeta_j^*} \vec{p}_j \vec{p}_i^{\,\dagger},
\nonumber \\
U_k^{\dagger} & = &  1 - i \sum_{n,m} (\widetilde{S_k^{-1}})_{nm} \frac{1}{\zeta_n} \vec{p}_m \vec{p}_n^{\,\dagger}.
\label{5.12}
\end{eqnarray}
where $S_k$ is the matrix obtained from $S$ by deleting the $k$-th row and column.
Then we find
\begin{equation}
U U_k^{\dagger} = 1 + \kappa_k \vec{Q}_k \vec{Q}_k^{\,\dagger}, \quad \kappa_k = \frac{(\zeta_k-\zeta_k^*)}{\zeta_k^*} \quad ({\rm no\ }k {\rm -sum})
\label{5.13}
\end{equation}
with
\begin{eqnarray}
\vec{Q}_k & = & \vec{K}_k (\vec{K}_k^{\,\dagger}\vec{K}_k)^{-1/2},
\nonumber \\
\vec{K}_k  &  = &   \sum_j (S^{-1})_{kj} \frac{1}{\zeta_j^*} \vec{p}_j. 
\label{5.14}
\end{eqnarray}
The normalization factor entering (\ref{5.14}) is given by
\begin{equation}
\vec{K}_k^{\,\dagger} \vec{K}_k  =  -i \left(\frac{1}{\zeta_k^*}-\frac{1}{\zeta_k} \right) (S^{-1})_{kk} \quad ({\rm no\ }k {\rm -sum}).
\label{5.15}
\end{equation}
These results will be used again in the applications in Sects.~\ref{sect8} and \ref{sect9}.

\section{General case including breathers}
\label{sect6}

In the general case, the matrix $\omega$ is non-diagonal. This describes breathers, solitons that are time dependent in their rest frame. The simplest
breather requires a 2$\times$2 block submatrix in $\omega$. Additional diagonal elements then describe solitons in interaction with the breather and each other. 
Larger block submatrices would correspond to more complex breathers built out of more than two twisted kink constituents. In the preceding chapter, an important step
was going from the diagonal matrix $v$ to the  inverse diagonal matrix $V$. This enabled us to exhibit the flavor structure and the possible intermediate
vacua in a clear fashion. Here we generalize this procedure to non-diagonal matrices $v,V$.

Recall that 
\begin{equation}
v= E^{-1} \omega (E^{\dagger})^{-1}, \quad v_{ij} = \frac{\omega_{ij}}{e_i e_j^*}
\label{6.1}
\end{equation}
and introduce the inverse of the matrix $v$ as
\begin{equation}
V=v^{-1}= E^{\dagger} \omega^{-1} E, \quad V_{ij} = e_i^* (\omega^{-1})_{ij} e_j.
\label{6.2}
\end{equation}
Our starting point is the identity for the inverse of a sum of two matrices (see Sect.~\ref{sect4}),
\begin{eqnarray}
\frac{1}{v+S} & = & \frac{{\rm adj}(v+S)}{\det (v+S)},
\nonumber \\
{\rm adj}(v+S) & = & {\rm adj\,}S + \sum_{r=1}^{N-1} \sum_{\alpha,\beta} (-1)^{s(\alpha)+ s(\beta)} \det v[\alpha|\beta] \widetilde{{\rm adj\,} S(\alpha|\beta)} ,
\nonumber \\
\det (v+S) & = & \det v + \det S + \sum_{r=1}^{N-1} \sum_{\alpha,\beta} (-1)^{s(\alpha)+ s(\beta)} \det v[\alpha|\beta] \det S(\alpha|\beta) .
\label{6.3}
\end{eqnarray}
Using Jacobi's complementary minor formula, we can express $\det v[\alpha|\beta]$ by the determinant of a submatrix of $V=v^{-1}$ as follows
\begin{equation}
\det v[\alpha|\beta| = (-1)^{s(\alpha)+s(\beta)} \frac{\det V(\beta|\alpha)}{\det V}.
\label{6.4}
\end{equation}
Replacing $\det v$ by $1/\det V$ and expanding numerator and denominator in (\ref{6.3}) by $\det V$,  we get
\begin{equation}
\frac{1}{V^{-1}+S} = \frac{ \sum_{r=1}^{N-1} \sum_{\alpha,\beta} \det V(\beta|\alpha) \widetilde{{\rm adj\,} S(\alpha|\beta)} + \det V {\rm adj\,}S}{1 + 
\sum_{r=1}^{N-1} \sum_{\alpha,\beta} \det V(\beta|\alpha) \det S(\alpha|\beta)+ \det V \det S}.
\label{6.5}
\end{equation}
As in the diagonal case, it is more convenient to switch notation from $V(\beta|\alpha)$ where rows $\beta$ and columns $\alpha$ are missing to $V[\beta'|\alpha']$
where the complementary rows $\beta'$ and complementary columns $\alpha'$ are kept. Since both $S$ and $V$ in (\ref{6.5}) now involve round brackets, we can switch both matrices to square brackets
and sum over the complementary sequences $\alpha',\beta'$
\begin{equation}
V(\beta|\alpha) = V[\beta'|\alpha'], \quad S(\alpha|\beta) = S[\alpha'|\beta'].
\label{6.6}
\end{equation}
The range of $r$ from 1 to $N-1$ does not change under this transition.
We also have to redefine the tilde-symbol accordingly: $\widetilde{ {\rm adj\,} S(\alpha|\beta)}$ meant that one has to fill the rows $\beta$ and columns $\alpha$ 
with zero's. Correspondingly,  $\widetilde{ {\rm adj\,} S[\alpha'|\beta']}$ now instructs us to fill the rows complementary to $\beta'$ and the columns
complementary to $\alpha'$ with 0's. 
At the end we rename the summation indices $\alpha',\beta'$ into $\alpha,\beta$ to ease the notation and find
\begin{equation}
\frac{1}{V^{-1}+S} = \frac{ \sum_{r=1}^{N} \sum_{\alpha,\beta} \det V[\beta|\alpha] \widetilde{{\rm adj\,} S[\alpha|\beta]}}{1 + 
\sum_{r=1}^{N} \sum_{\alpha,\beta} \det V[\beta|\alpha] \det S[\alpha|\beta]}.
\label{6.7}
\end{equation}
This is the generalization of (\ref{5.3}) to non-diagonal $V$. We now insert this expression into $\Delta$, Eq.~(\ref{4.6}), 
treating the $r=N$ terms in the sums separately for notational reasons, with the result
\begin{eqnarray}
\Delta & = &  \frac{\cal N}{\cal D},
\nonumber \\
{\cal N} & = & 1  + \sum_{r=1}^{N-1} \sum_{\alpha,\beta} \xi_{\alpha \beta}  U_{\alpha \beta} \det V[\beta|\alpha] + \xi  U \det V,
\nonumber \\
{\cal D} & = & 1 + \sum_{r=1}^{N-1} \sum_{\alpha,\beta} \xi_{\alpha \beta} \det V[\beta|\alpha] + \xi \det V.
\label{6.8}
\end{eqnarray}
Generalizing the $\xi_{\alpha}$ in Sect.~\ref{sect5} we have defined
\begin{equation}
\xi_{\alpha \beta} = \det S[\alpha|\beta|, \quad \xi = \det S.
\label{6.9}
\end{equation}
The twist matrices $U_{\alpha \beta}$ in flavor space are the generalization of $U_{\alpha}$ in Eq.~(\ref{5.5}),
\begin{eqnarray}
U_{\alpha \beta} & = & 1 + i \sum_{i,j} \left(\widetilde{S[\alpha|\beta]^{-1}}\right)_{ij}\frac{1}{\zeta_j^*} \vec{p}_j \vec{p}_i^{\,\dagger},
\nonumber \\
 U & = & 1 + i \sum_{i,j} (S^{-1})_{ij} \frac{1}{\zeta_j^*} \vec{p}_j \vec{p}_i^{\,\dagger}.
\label{6.10}
\end{eqnarray}
The matrix $U$ in the last line is the asymptotic vacuum at $x \to \infty$, independently of whether $\omega$ is diagonal or non-diagonal. Unlike the $U_{\alpha}$ in the preceding section, 
the $U_{\alpha \beta}$ are no longer unitary but satisfy the generalised unitarity relation
\begin{equation}
U_{\alpha \beta} U_{\beta \alpha}^{\dagger} = 1.
\label{6.11}
\end{equation}
The proof is similar to the proof that $U U^{\dagger}=1$ following Eq.~(\ref{5.6}). Start from
\begin{eqnarray}
U_{\alpha \beta} & = & 1+ i \sum_{n,j}  \left(\widetilde{S[\alpha|\beta]^{-1}}\right)_{nj}\frac{1}{\zeta_j^*} \vec{p}_j \vec{p}_n^{\,\dagger},
\nonumber \\
U_{\beta \alpha}^{\dagger} & = & 1- i \sum_{i,m}  \left(\widetilde{S[\beta|\alpha]^{-1}}\right)_{mi}^*\frac{1}{\zeta_i} \vec{p}_m \vec{p}_i^{\,\dagger}.
\label{6.12}
\end{eqnarray}
Notice that with this choice of dummy indices, we must have $n\in \beta, j\in \alpha, i \in \beta, m \in \alpha$ due to the 
definition of the tilde symbol. The product $U_{\alpha \beta} U_{\beta \alpha}^{\dagger}$ yields 1, two terms linear in the dyadics and a term quadratic in the dyadics. 
This last term reads
\begin{equation}
\left. U_{\alpha \beta} U_{\beta \alpha}^{\dagger} \right|_{\rm quad}  =  \sum_{i,j} \frac{1}{\zeta_i \zeta_j^*} \vec{p}_j \vec{p}_i^{\,\dagger} 
\sum_{n,m} \sigma_{nm}  \left(\widetilde{S[\alpha|\beta]^{-1}}\right)_{nj}  \left(\widetilde{S[\beta|\alpha]^{-1}}\right)_{mi}^*
\label{6.13}
\end{equation}
where 
\begin{equation}
\sigma_{nm} = \vec{p}_n^{\,\dagger} \vec{p}_m =  -i (\zeta_n-\zeta_m^*) S_{mn}.
\label{6.14}
\end{equation}
In the term containing $\zeta_m^*$, we perform the summation over $n$ as follows
\begin{equation}
\sum_n  S_{mn} (\widetilde{S[\alpha|\beta]^{-1}})_{nj} = \delta_{mj}.
\label{6.15}
\end{equation}
This term cancels the term linear in the dyadics contained in $U_{\beta \alpha}^{\dagger}$. In the term containing $\zeta_n$, we use again (\ref{6.14})
together with the hermiticity of $S$ to get
\begin{equation}
\sum_m S_{nm}^*  (\widetilde{S[\beta|\alpha]^{-1}})_{mi}^* = \delta_{ni}.
\label{6.16}
\end{equation} 
The resulting term cancels the term linear in the dyadics contained in $U_{\alpha \beta}$. This proves the assertion (\ref{6.11}).
Finally, products of $V_i$'s in the diagonal case (\ref{5.5})
are replaced by determinants of submatrices of $V$ depending only on the $e_i,e_i^*$ and $\omega_{ij}$.

This is the end of the formal part of the present work. The first step has been to transform the original expressions (\ref{3.10}) for the mean field and (\ref{3.5},\ref{3.6},\ref{3.9}) for the spinors
into (\ref{4.6}) and (\ref{4.7},\ref{4.8}) where the flavor structure has been exposed. A central element of all these expressions is the inverse matrix $(v+S)^{-1}$.
In order to separate space-time dependence from the other dependencies, we derived a closed expression for the inverse of a sum of two
square matrices, starting from a well-known formula for the determinant of a sum of matrices. The final result for the mean field $\Delta$ is surprisingly simple and given in
(\ref{5.4},\ref{5.5}) for solitons only (diagonal $\omega$) and in (\ref{6.8},\ref{6.9},\ref{6.10}) for the general case including breathers (non-diagonal $\omega$). In the following sections,
we shall use these results as starting point to illustrate the formalism with simple examples. 

\section{Example I: Single twisted kink}
\label{sect7}

We first have to understand thoroughly a single twisted kink, the basic building block of all more complex configurations.
For one soliton, $v+S$ is a 1$\times$1 ``matrix" so that there is no issue of matrix inversion.  
Nevertheless, the 
mean field $\Delta$ is a $N_f \times N_f$ matrix and we focus on the role of the flavor vector $\vec{p}_1$. 
We start from 
\begin{equation}
\left(\frac{1}{v+S}\right)_{11} = \frac{V_1}{1+S_{11} V_1}, \quad S_{11}= \frac{i}{\zeta_1-\zeta_1^*}.
\label{7.1}
\end{equation}
Upon defining 
\begin{equation}
\tilde{V_1} = S_{11} V_1,
\label{7.2}
\end{equation}
we find a very simple expression for $\Delta$,
\begin{equation}
\Delta  =   \frac{1+ U_1 \tilde{V}_1}{1+\tilde{V}_1}
\label{7.3}
\end{equation}
with
\begin{equation}
U_1 = 1 + \kappa_1 \vec{p}_1  \vec{p}_1^{\,\dagger}, \quad \kappa_1 = \frac{\zeta_1-\zeta_1^*}{\zeta_1^*}
\label{7.4}
\end{equation}
We recall from the one-flavor model that
\begin{equation}
\zeta_1 = - \frac{ e^{-i\theta_1}}{\eta_1}, \quad \eta_1 = \sqrt{\frac{1+v_1}{1-v_1}},
\label{7.5}
\end{equation}
where $v_1$ is the velocity of the soliton. The function $\tilde{V}_1$ is independent of $N_f$ and given by
\begin{equation}
\tilde{V}_1 = \frac{S_{11}}{\omega_{11}}e^{2 \sin \theta_1 x'}
\label{7.6}
\end{equation}
with $x'$ the boosted form of $x$,
\begin{equation}
x' = \frac{x-v_1t}{\sqrt{1-v_1^2}}.
\label{7.7}
\end{equation}
The twisted kink interpolates between the vacua $\Delta=1$ at $x\to - \infty$ and $\Delta=U_1$ at $x\to \infty$, the twist matrix $U_1$ replacing the phase factor $e^{-2i\theta_1}$ in the one-flavor case,
\begin{equation}
U_1 = 1 + (e^{-2i\theta_1}-1)\vec{p}_1 \vec{p}_1^{\,\dagger} = \exp \left\{-2i\theta_1 \vec{p}_1 \vec{p}_1^{\,\dagger} \right\}.
\label{7.8}
\end{equation}
Equivalently, $U_1$ may be written in the canonical form of a U($N_f$) group element
\begin{equation}
U_1 = \exp \left\{ -2i \theta_1 n^a \lambda^a \right\} , \quad n^a = \vec{p}_1^{\,\dagger} \lambda^a \vec{p}_1, \quad n^a n^a = 1.
\label{7.9}
\end{equation}
Thus the twist angle has the same interpretation as in the one-flavor case, namely as a ``rotation angle". The novel vectors $\vec{p}_1$
(in the fundamental representation) serve to define the ``rotation axis" $n^a$, a unit vector in the adjoint representation.
If there is only one soliton, we are free to choose a frame in which this axis is pointing into the 0-direction. Then  
everything reduces to the U(1) case.

Due to the close relationship with the single flavor case, we refrain from discussing the continuum spinors. It is of some interest 
though to look at the normalized bound state spinors with Dirac components 
\begin{equation}
\vec{\varphi}_{1,1}   =  - \zeta_1^*  \vec{\varphi}_{2,1}=   \frac{1}{\sqrt{2} S_{11}} \frac{1}{e_1^*} \frac{\tilde{V}_1}{1+\tilde{V}_1} \vec{p}_1.
\label{7.10}
\end{equation}
In the one flavor case, the total fermion charge of the bound state is given by the occupation fraction,
\begin{equation}
Q = \nu_1 = \frac{\theta_1}{\pi},
\label{7.11}
\end{equation}
where we have used the self-consistency condition. The analogous calculation in the multiflavor case yields the flavor charge $Q^a = \langle \vec{\varphi}^{\dagger} \lambda^a \vec{\varphi} \rangle$
in the bound state to be 
\begin{equation}
Q^a = \nu_1 \vec{p}_1^{\,\dagger} \lambda^a \vec{p}_1 = \frac{\theta_1}{\pi}  n^a.
\label{7.12}
\end{equation}
This gives yet another physical interpretation of $\vec{p}_1$, namely determining the direction of the flavor vector (a generalization of the isospin vector) associated with the 
bound orbit. We have already mentioned
that the full charge density vanishes due to a cancellation between continuum and bound states, a consequence of chiral symmetry and current conservation.
In the one flavor case, it was recently pointed out that the total charge of a twisted kink  is infrared sensitive and needs some regularization, either by a
small bare fermion mass \cite{18} or a finite box \cite{19}. The conclusion was that the charge is spread out over the whole space in the thermodynamic and chiral limit and hence becomes invisible.
Nevertheless, a regularized integrated charge can be defined consistently and agrees with the charge of the fermions in the bound state. For a single 
soliton, the same arguments could be applied here as well, giving a more direct physical meaning to the vector $n^a$ as flavor vector of the twisted kink as a whole.

\section{Example II: Scattering and bound state of two twisted kinks}
\label{sect8}

Eqs.~(\ref{5.4},\ref{5.5}) yield the following mean field $\Delta$ for two twisted kinks,
\begin{equation}
\Delta = \frac{1+ \xi_1 V_1 U_1 + \xi_2 V_2 U_2 + \xi_{12} V_1V_2 U_{12}}{1+ \xi_1 V_1 + \xi_2 V_2 + \xi_{12} V_1 V_2}.
\label{8.1}
\end{equation}
Introducing
\begin{equation}
\tilde{V}_i = \xi_i V_i, \quad \tilde{\xi}_{12} = \frac{\xi_{12}}{\xi_1 \xi_2},
\label{8.2}
\end{equation}
this goes over into
\begin{equation}
\Delta = \frac{1 + \tilde{V}_1 U_1 + \tilde{V}_2 U_2 +  \tilde{\xi}_{12} \tilde{V}_1 \tilde{V}_2  U_{12}}{1 + \tilde{V}_1 + \tilde{V}_2 + \tilde{\xi}_{12} \tilde{V}_1 \tilde{V}_2}.
\label{8.3}
\end{equation}
Using Eqs.~(\ref{4.3},\ref{5.5}), we evaluate $\tilde{\xi}_{12}$,
\begin{eqnarray}
\xi_i & = & S_{ii} = \frac{i}{\zeta_i - \zeta_i^*},
\nonumber \\
\tilde{\xi}_{12} & = & 1 - \frac{(\zeta_1-\zeta_1^*)(\zeta_2-\zeta_2^*)}{(\zeta_2-\zeta_1^*)(\zeta_1-\zeta_2^*)} \sigma_{12} \sigma_{21}.
\label{8.4}
\end{eqnarray}
The twist factors entering the numerator of $\Delta$ are 
\begin{eqnarray}
U_i & = & 1+ \kappa_i \vec{p}_i \vec{p}_i^{\,\dagger}, \quad \kappa_i = \frac{\zeta_i-\zeta_i^*}{\zeta_i^*},
\nonumber \\
U_{12} & =  & 1 - \frac{\vec{p}_1 \vec{p}_1^{\,\dagger}}{(\zeta_2-\zeta_2^*)\zeta_1^* \det S} - \frac{\vec{p}_2 \vec{p}_2^{\,\dagger}}{(\zeta_1-\zeta_1^*)\zeta_2^* \det S}
\nonumber \\ 
& & + \frac{\sigma_{12} \vec{p}_1 \vec{p}_2^{\,\dagger}}{(\zeta_1-\zeta_2^*)\zeta_1^* \det S} + \frac{\sigma_{21}\vec{p}_2 \vec{p}_1^{\,\dagger}}{(\zeta_2-\zeta_1^*)\zeta_2^* \det S},
\nonumber \\
\det S & = & - \frac{1}{(\zeta_1-\zeta_1^*)(\zeta_2-\zeta_2^*)} \tilde{\xi}_{12}.
\label{8.5}
\end{eqnarray}
The leftmost solitons (incoming kink I or outgoing kink II) are characterized by intrinsic flavor vectors $\vec{p}_1, \vec{p}_2$. The rightmost solitons (outgoing kink I  or  incoming kink II) have the intrinsic twist matrices
\begin{eqnarray}
U_{12} U_2^{\dagger} & = &  1 + \kappa_1 \vec{Q}_1 \vec{Q}_1^{\,\dagger},
\nonumber \\
U_{12} U_1^{\dagger} & = &  1 + \kappa_2 \vec{Q}_2 \vec{Q}_2^{\,\dagger},
\label{8.6}
\end{eqnarray}
with the intrinsic flavor vectors
\begin{eqnarray}
\vec{Q}_1 & = & {\cal N}_1\left[ \zeta_2^* (\zeta_2-\zeta_1^*) \vec{p}_1 - \zeta_1^* (\zeta_2-\zeta_2^*) \sigma_{21} \vec{p}_2 \right],
\nonumber \\
\vec{Q}_2 & = &  {\cal N}_2 \left[ \zeta_1^* (\zeta_1-\zeta_2^*) \vec{p}_2 - \zeta_2^* (\zeta_1-\zeta_1^*) \sigma_{12} \vec{p}_1 \right],
\nonumber \\
{\cal N}_1^{-2} & = &  \frac{\zeta_2 \zeta_2^*}{\zeta_1 \zeta_1^*} {\cal N}_2^{-2} =   (\zeta_1-\zeta_2^*)(\zeta_2-\zeta_1^*)(\zeta_1-\zeta_1^*)(\zeta_2-\zeta_2^*) \zeta_2 \zeta_2^* \det S .
\label{8.7}
\end{eqnarray}
Twisted kinks are now characterized by a twist angle and a twist axis. In the one-flavor case, the axis is frozen and the angle is conserved during the collision. The only observable 
of a two-soliton scattering event is then the time delay. For many flavors, the twist angles are still conserved, but the twist axes are rotated during the collision. There are now two
observables, the time delay and the change of orientation of the twist axis. Both of these depend on the relative orientation of the two flavor axes of the colliding solitons.
If we choose $\vec{p}_1=\vec{p}_2$, the flavor vectors are aligned and everything is concentrated in a single flavor component. Then we are back at the 
one-flavor case, including the time delay. In the other extreme, choosing $\vec{p}_1, \vec{p}_2$ or, equivalently, the flavor vectors $n_1^a,n_2^a$ to be orthogonal,  the two kinks decouple.
There is no scattering at all and the solitons cross each other without interaction. The novel feature of the multiflavor model is the fact that we can control the strength of the 
interaction between two solitons with the help of the parameter $\sigma_{12}=\vec{p}_1^{\,\dagger}\vec{p}_2$. To see how the theory interpolates between the extreme cases
$\sigma_{12}=1$ and $\sigma_{12}=0$ just discussed, we determine how the scattering observables depend on this parameter.
To this end, we first extract the asymptotic form of the incoming and outgoing solitons from the full expression for $\Delta$, Eq.~(\ref{8.3}),
\begin{eqnarray}
\Delta^{I}_{\rm in} & = & \lim_{\tilde{V}_2 \to 0} \Delta = \frac{1+ \tilde{V}_1 U_1}{1+\tilde{V}_1},
\nonumber \\
\Delta^{I}_{\rm out} & = & \lim_{\tilde{V}_2 \to \infty} \Delta = \left( \frac{1+ \tilde{\xi}_{12} \tilde{V}_1 U_{12} U_2^{\dagger}}{1 + \tilde{\xi}_{12}\tilde{V}_1} \right) U_2,
\nonumber \\
\Delta^{II}_{\rm in} & = & \lim_{\tilde{V}_1 \to \infty} \Delta = \left( \frac{1+ \tilde{\xi}_{12} \tilde{V}_2 U_{12} U_1^{\dagger}}{1 + \tilde{\xi}_{12}\tilde{V}_2} \right) U_1,
\nonumber \\
\Delta^{II}_{\rm out} & = & \lim_{\tilde{V}_1 \to 0} \Delta = \frac{1+ \tilde{V}_2 U_2}{1+\tilde{V}_2}.
\label{8.8}
\end{eqnarray}
In the 2nd and 3rd line we have exhibited the intrinsic form of the twisted kinks.
During the collision, the intrinsic flavor vector of soliton I changes from $\vec{p}_1$ to $\vec{Q}_1$, that of soliton II from $\vec{Q}_2$ to $\vec{p}_2$.
The change in flavor direction is characterized by 
\begin{equation}
(\vec{p}_i^{\,\dagger} \lambda^a \vec{p}_i)(\vec{Q}_i^{\,\dagger} \lambda^a \vec{Q}_i)= |\vec{p}_i^{\, \dagger}\vec{Q}_i|^2, \quad (i=1,2).
\label{8.9}
\end{equation}
The time delay can be found by equating
\begin{equation}
V_i(x,t-(\Delta t)_i)= {\tilde{\xi}_{12}} V_i(x,t).
\label{8.10}
\end{equation}
Consider two solitons with twist angles $\theta_1,\theta_2$ and equal and opposite velocity $\pm v$.
Introducing the parameter $\lambda=|\sigma_{12}|^2$, we find the time delay
\begin{eqnarray}
\sin \theta_1 (\Delta t)_1  =  - \sin \theta_2 (\Delta t)_2 & = &  \frac{\sqrt{1-v^2}}{2v}\ln \frac{1+v^2- (1-v^2)[ \lambda C_- + (1-\lambda)C_+]}{1+v^2 -C_+(1-v^2)} 
\nonumber \\
& \approx &  \frac{\sqrt{1-v^2}}{2v} \frac{(1-v^2)(C_+-C_-)}{1+v^2-(1-v^2)C_+} \lambda + {\rm O}(\lambda^2)
\label{8.11}
\end{eqnarray}
where
\begin{equation}
C_{\pm} = \cos (\theta_1 \pm \theta_2).
\label{8.12}
\end{equation}
The change in intrinsic flavor spin orientation of soliton I is 
\begin{eqnarray}
|\vec{p}_1^{\, \dagger} \vec{Q}_1|^2 -1 & = & - \frac{2 \sin^2 \theta_2 \lambda (1-\lambda) (1-v)^2}{1+v^2+(1-v^2)[C_+ (1-\lambda)+C_-\lambda]}
\nonumber \\
& \approx & - \frac{2 \sin^2 \theta_2 \lambda  (1-v)^2}{1+v^2+(1-v^2)C_+} + {\rm O}(\lambda^2) .
\label{8.13}
\end{eqnarray}
For soliton II, we get the same result except for the substitution $v \to -v$. 

Eqs.~(\ref{8.11}-\ref{8.13}) confirm that $\lambda$ governs the interaction strength between the two solitons, here exhibited in the observables time delay and flavor spin rotation.
As expected, the scattering observables vanish 
in the limit $\lambda \to 0$. In the opposite limit $\lambda \to 1$, the orientation of the axis is not changed and the time delay reduces to what is known from the one-flavor 
model.

Finally, we mention that by choosing $v=0$ in the above expressions, we can specialize the scattering problem to the bound state of two
twisted kinks. The parameter $\lambda=|\sigma_{12}|^2$ again allows us to interpolate between the one-flavor bound state ($\lambda=1$) and a pair of
non interacting single twisted kinks ($\lambda=0$). 

\section{Example III: Breathers}
\label{sect9}

Finally, we turn to the twisted breather. A breather at rest can be generated by choosing $\eta_1=\eta_2=1$ and a non-diagonal matrix $\omega$.
An example for a two-soliton breather has been discussed in the case $N_f=2$ before \cite{14}, and we have reproduced these results to test our present formalism
for non-diagonal matrix $\omega$. Up to translations in space and time, $\omega$ can be chosen as
\begin{equation}
\omega = \left( \begin{array}{cc} \sec \chi & \tan \chi  \\ \tan \chi & \sec \chi  \end{array} \right),  \quad \det \omega=1 .
\label{9.1}
\end{equation}
The formalism of Sect.~\ref{sect6} then yields the following expression for the mean field $\Delta$
\begin{eqnarray}
\Delta & = & \frac{{\cal N}_{\rm b}}{{\cal D}_{\rm b}}
\nonumber \\
{\cal N}_{\rm b} &  = &  1 + \sec \chi (\xi_{1,1}U_{1,1} V_1+ \xi_{2,2} U_{2,2} V_2) - \tan \chi (\xi_{1,2} U_{1,2} K_{21} + \xi_{2,1} U_{2,1} K_{12}) + \xi_{12,12} U_{12,12} V_1 V_2
\nonumber \\
{\cal D}_{\rm b} & = & 1+    \sec \chi (\xi_{1,1} V_1+ \xi_{2,2}  V_2) - \tan \chi (\xi_{1,2}  K_{21} + \xi_{2,1}  K_{12}) + \xi_{12,12}V_1 V_2
\label{9.2}
\end{eqnarray}
with 
\begin{equation}
V_i = |e_i|^2, \quad  K_{ij} = e_i^*e_j.
\label{9.3}
\end{equation}
The characteristic novel feature of the breather are the $K_{ij}$ yielding oscillations with the frequency $\Omega = \cos \theta_1 - \cos \theta_2$. The $\xi_{\alpha \beta}$ and $U_{\alpha \beta}$
can now easily be constructed from submatrices of $S$, see Eqs.~(\ref{6.9},\ref{6.10}). In this way we recover the results from Ref.~\cite{14}.
If we choose $\vec{p}_1$ and $\vec{p}_2$ to be parallel, we come back to the known one-flavor breather.  
Choosing $\vec{p}_1$ and $\vec{p}_2$ to be orthogonal one finds that
the diagonal components $\Delta_{11}, \Delta_{22}$ are static, whereas the off-diagonal components $\Delta_{12}, \Delta_{21}$ oscillate
with the same frequency as the one-flavor breather.
For any other choice of the angle between $\vec{p}_1,\vec{p}_2$, all components of $\Delta$ start to oscillate with the same frequency but different phases.
Thus, unlike in the soliton case, the interaction does not disappear if the flavor vectors are orthogonal. It is mediated by the off-diagonal matrix elements of $\omega$.

The formalism enables us to go beyond this type of complexity. By way of example, we sketch how one would analyse scattering between
a breather and a soliton. This shows that one gets some analytical insight even without carrying out the full, tedious calculation to the end. We  focus on the 
asymptotics of the scattering event to get an idea how the breather and the soliton are affected by a collision.

To this end, we consider a three soliton configuration with the $\omega$ matrix 
\begin{equation}
\omega = \left( \begin{array}{ccc} \sec \chi & \tan \chi & 0 \\ \tan \chi & \sec \chi & 0  \\ 0 & 0 & 1 \end{array} \right),  \quad \det \omega=1 .
\label{9.4}
\end{equation}
Evaluating the determinants of submatrices of $V$, one finds that the formalism predicts already quite a number of terms to account for the breather, the soliton and their interaction,
\begin{eqnarray}
\Delta & = & \frac{\cal N}{\cal D},
\nonumber \\
{\cal N} & = & {\cal N}_{\rm b} + \xi_{3,3} U_{3,3}V_3 + \sec \chi (\xi_{13,13} U_{13,13}V_1+ \xi_{23,23}U_{23,23} V_2)V_3
\nonumber \\ 
& & - \tan \chi (\xi_{13,23} U_{13,23} K_{21} + \xi_{23,13} U_{23,13}K_{12})V_3 + \xi U V_1 V_2 V_3,
\nonumber \\
{\cal D}  & = & {\cal D}_{\rm b} +  \xi_{3,3}V_3 + \sec \chi (\xi_{13,13} V_1+ \xi_{23,23}  V_2)V_3 
\nonumber \\
& & - \tan \chi (\xi_{13,23}  K_{21} + \xi_{23,13} K_{12})V_3 + \xi  V_1 V_2 V_3.
\label{9.5}
\end{eqnarray}
We recognize the breather pieces ${\cal N}_{\rm b},{\cal D}_{\rm b}$ from Eq.~(\ref{9.2}) which we did not spell out again, terms $\sim V_3$ involving only the soliton and interaction terms. 
As is always the case, the denominator can be obtained from the numerator by setting all $U_{\alpha\beta}$ equal to 1. It would now be straightforward to evaluate all the coefficients and twist factors. However, it is perhaps
more instructive to extract the asymptotics. In the real world, this would contain all scattering observables. We have to treat $V_1, V_2, K_{12},K_{21}$ as being of the same order and consider
as in Eq.~(\ref{8.8})
\begin{eqnarray}
\Delta_{\rm b,in} & = & \lim_{V_3 \to 0} \Delta = \frac{{\cal N}_{\rm b}}{{\cal D}_{\rm b}},
\nonumber \\
\Delta_{\rm b,out} & = & \lim_{V_3 \to \infty} \Delta = \left( \frac{{\cal N}_{\rm b'}}{{\cal D}_{\rm b'}}\right) U_{3,3},
\nonumber \\
{\cal N}_{\rm b'} &  = &  1 + \sec \chi \left(\xi_{13,13}\xi_{3,3}^{-1} U_{13,13}U_{3,3}^{\dagger}  V_1+ \xi_{23,23}\xi_{3,3}^{-1} U_{23,23}U_{3,3}^{\dagger} V_2\right) 
\nonumber \\
& & - \tan \chi \left(\xi_{13,23}\xi_{3,3}^{-1} U_{13,23}U_{3,3}^{\dagger}  K_{21} + \xi_{23,13}\xi_{3,3}^{-1}  U_{23,13}U_{3,3}^{\dagger} K_{12}\right) + \xi \xi_{3,3}^{-1} UU_{3,3}^{\dagger} V_1 V_2,
\nonumber \\
{\cal D}_{\rm b'} & = & 1 + \sec \chi \left(\xi_{13,13}\xi_{3,3}^{-1} V_1+ \xi_{23,23}\xi_{3,3}^{-1} V_2\right) - \tan \chi \left(\xi_{13,23}\xi_{3,3}^{-1} K_{21} 
+ \xi_{23,13}\xi_{3,3}^{-1}  K_{12}\right) + \xi \xi_{3,3}^{-1}  V_1 V_2,
\nonumber \\
\Delta_{\rm s,in} & = & \lim_{V_1 \to \infty} \Delta = \left( \frac{1+ \xi \xi_{12,12}^{-1} V_3 U U_{12,12}^{\dagger}}{1+ \xi \xi_{12,12}^{-1} V_3}\right) U_{12,12},
\nonumber \\
\Delta_{\rm s,out} & = & \lim_{V_1 \to 0} \Delta = \frac{1+ \xi_{3,3} V_3 U_{3,3}}{1+\xi_{3,3} V_3}.
\label{9.6}
\end{eqnarray}
The subscripts b and s refer to breather and soliton, respectively. 

How are soliton and breather affected during the scattering event? For the soliton, we can again read off the time delay and the change in flavor orientation.
For the soliton-soliton collision discussed in Sec.~\ref{sect7}, the time delay was proportional to $\ln \tilde{\xi}_{12}$ where $\tilde{\xi}_{12}=\xi_{12}\xi_{1}^{-1}
\xi_2^{-1}$. Now the same formula applies except that the argument of the logarithm is $\xi \xi_{12,12}^{-1} \xi_{3,3}^{-1}$. The intrinsic flavor vector
changes from the vector defined by $U U_{12,12}^{\dagger}$ to $\vec{p}_3$. For the breather, things are more complicated since the different building blocks
are affected differently, so that the whole internal structure changes. For the time delays, the relevant arguments of the logarithms are
\begin{eqnarray}
V_1: & & \frac{\xi_{13,13}}{\xi_{1,1} \xi_{3,3}}, \quad V_2: \quad \frac{\xi_{23,23}}{\xi_{2,2} \xi_{3,3}},\quad V_1V_2: \quad \frac{\xi}{\xi_{12,12} \xi_{3,3}}
\nonumber \\
K_{21} : & & \frac{\xi_{13,23}}{\xi_{1,2}\xi_{3,3}}, \quad K_{12}: \quad \frac{\xi_{23,13}}{\xi_{2,1}\xi_{3,3}}
\label{9.7}
\end{eqnarray}
All the $\xi_{\alpha \beta}$ can be evaluated as determinants of submatrices of $S$. More interesting is perhaps the observation that the results are so regular that one
can guess the general principle behind them, even for more complicated collisions. Since different parts of the breather suffer different time delays, the 
structure changes. This is also true for the changes in internal flavor direction which are also different for different components of the breather.

\section{Summary and conclusions}
\label{sect10}

The chiral GN model with ${\rm U}_L(N_f)\times{\rm U}_R(N_f)$ symmetry is probably one of the most complicated quantum field theories that one can still solve
analytically, at least in the large $N_c$ limit where semiclassical methods become exact. This has incited us to reconsider the question of soliton dynamics
in this model. We were able to build our study on existing results from both condensed matter and particle theory. Although the most important steps in
solving the problem had already been done, we felt that the analytical results for the interaction of solitons and breathers had not yet been 
cast into a sufficiently intuitive form. In the present work, we have therefore reformulated the general solution. The main idea was to manipulate the formal expression
for the mean field in such a way that the ($x,t$) dependent factors are neatly separated from constant coefficients and twist matrices. 
Explicit expressions for the coefficients and twist matrices could be derived. We were guided by previous results for scattering of $N$
solitons in the single flavor model, simpler than the general expressions. At the core of this problem was the necessity to invert
a sum of two square matrices with dimension given by the number of solitons involved. Starting from a well-known formula for the determinant of a sum 
of two matrices, we derived the corresponding formula for the inverse of a sum and applied it to the problem at hand. If one disregards breathers, the 
result is particularly simple. The mean field is represented as a quotient of two multivariate polynomials in the basic
exponential soliton functions $V_i$, much like in the one flavor case. The novel feature as compared to a single flavor is the fact that the twist factors
in the numerator now become unitary $N_f \times N_f$ matrices, as opposed to phase factors (or U(1) elements) before. Both the coefficients
and the twist matrices are given by compact, simple formulas involving submatrices of a constant matrix $S$. This procedure can be extended 
in the presence of breathers where things get more involved. Here, submatrices of $S$ appear where different rows and columns are kept.
The $V_i$ have to be replaced by determinants of block matrices characteristic for breathers. The example of breather-soliton scattering 
has been used to illustrate the advantage of the new formulation, for instance in extracting the asymptotics of the scattering process in a simple
manner.

As an outlook, we would like to come back to the formula for inverting a sum of two matrices. From a pragmatic point of view, if one is only interested in
the (numerical or analytical) result, this is of little help. The expression on the right hand side of Eq.~(\ref{4.16}) is obviously more complicated than the
original problem. However we have seen that it has merits for organizing the analytical result, disentangling the separate contributions from the two matrices
in the inverse sum. In our application, everything seems to fall into place. It will be interesting to see whether this has other applications in physics as well.

\section*{Acknowledgement}
The author would like to thank Oliver Schnetz for his advice concerning the proof of Eq.~(\ref{4.16}). 

\section*{APPENDIX: DERIVATION OF EQS.~(\ref{5.13}-\ref{5.15})}

$U=U_{1..N}$ denotes the unitary matrix with the largest number of indices. $U_k$ is the unitary matrix for the $N-1$ soliton term where the index $k$ 
is missing. We want to derive an expression for $U U_k^{\dagger}$, an intrinsic soliton. Our starting point is
\begin{eqnarray}
U & = & 1 + i \sum_{i,j} (S^{-1})_{ij} \frac{1}{\zeta_j^*} \vec{p}_j \vec{p}_i^{\,\dagger},
\nonumber \\
U_k^{\dagger} & = &  1 - i \sum_{n,m} \left(\widetilde{S_k^{-1}}\right)_{nm} \frac{1}{\zeta_n} \vec{p}_m \vec{p}_n^{\,\dagger}.
\label{A1}
\end{eqnarray}
We evaluate
\begin{eqnarray}
U U_k^{\dagger}-1 & = &  (U-1) + (U_k^{\dagger}-1) + (U-1)(U_k^{\dagger}-1)
\nonumber \\
& = & {\cal A}_1 + {\cal A}_2 + {\cal A}_3.
\label{A2}
\end{eqnarray}
We only need to compute  ${\cal A}_3$, 
\begin{eqnarray}
{\cal A}_3 & = & \sum_{j,n} Y_{jn}^k \frac{1}{\zeta_j^* \zeta_n} \vec{p}_j \vec{p}_n^{\,\dagger},
\nonumber \\
Y_{jn}^k & = & \sum_{i,m} (S^{-1})_{ij} \left(\widetilde{S^{-1}_k }\right)_{nm} \sigma_{im}.
\label{A3}
\end{eqnarray}
Expressing $\sigma_{im}$ through $S_{mi}$, find
\begin{eqnarray}
Y_{jn}^k & = & -i \sum_i \zeta_i (S^{-1})_{ij} X_{ni} ^k+ i \sum_m \zeta_m^* \left({\widetilde{S}_k^{-1}}\right)_{nm} \tilde{X}_{mj},
\nonumber \\
X_{ni}^k & = & \sum_m \left( \widetilde{S_k^{-1}}\right)_{nm} S_{mi},
\nonumber \\
\tilde{X}_{mj} & = & \sum_i S_{mi} (S^{-1})_{ij},
\label{A4}
\end{eqnarray}
where
\begin{eqnarray}
X_{ni}^k & = & \delta_{ni} \quad {\rm for\ } n\neq k, i \neq k,
\nonumber \\
X_{ni}^k & = & 0 \quad {\rm for \ }n=k,
\nonumber \\
X_{nk}^k & = & - \frac{({\rm adj\,} S)_{nk}}{\det( S_k) },
\nonumber \\
\tilde{X}_{mj} & = & \delta_{mj}.
\label{A5}
\end{eqnarray}
This yields
\begin{equation}
Y_{jn} = (1-\delta_{nk}) \left[ i \zeta_j^* \left(\widetilde{S_k^{-1}}\right)_{nj} - i \zeta_n (S^{-1})_{nj} + i \zeta_k \frac{(S^{-1})_{kj} ({\rm adj\,} S)_{nk}}{\det (S_k)}\right].
\label{A6}
\end{equation}
Insert this result into ${\cal A}_3$, Eq.~(\ref{A3}). The first term cancels ${\cal A}_2$. The second term partially cancels ${\cal A}_1$, leaving the $n=k$ term.
This combines with the 3rd term to the simple final result
\begin{equation}
U U_k^{\dagger}-1  = i \sum_{j,n} \frac{(S^{-1})_{kj} (S^{-1})_{nk} \det S}{\det S_k} \frac{1}{\zeta_j^* \zeta_n} \vec{p}_j \vec{p}_n^{\,\dagger}.
\label{A7}
\end{equation}
This factorizes indeed,
\begin{eqnarray}
U U_k^{\dagger}-1  & = & i \frac{\zeta_k \det S}{\det S_k} \vec{K}_k \vec{K}_k^{\,\dagger},
\nonumber \\
\vec{K}_k & = & \sum_j (S^{-1})_{kj} \frac{1}{\zeta_j^*} \vec{p}_j ,
\nonumber \\
\vec{K}_k^{\,\dagger} & = & \sum_i (S^{-1})_{ik} \frac{1}{\zeta_i} \vec{p}_i^{\,\dagger}.
\label{A8}
\end{eqnarray}
The norm is
\begin{eqnarray}
\vec{K}_k^{\,\dagger} \vec{K}_k & = & \sum_{i,j} \frac{1}{\zeta_i \zeta_j^*} (S^{-1})_{ik} (S^{-1})_{kj} \sigma_{ij}
\nonumber \\
& = & -i \left(\frac{1}{\zeta_k^*}-\frac{1}{\zeta_k} \right) (S^{-1})_{kk} \quad ({\rm no\ }k {\rm -sum}).
\label{A9}
\end{eqnarray}
Using ($S^{-1})_{kk} = \det S_k/\det S$ and introducing normalized vectors $\vec{Q}_k=\vec{K}_k (\vec{K}_k^{\,\dagger}\vec{K}_k)^{-1/2}$, we finally arrive
at the standard form 
\begin{equation}
U U_k^{\dagger} = 1 + \kappa_k \vec{Q}_k \vec{Q}_k^{\,\dagger}, \quad \kappa_k = \frac{(\zeta_k-\zeta_k^*)}{\zeta_k^*}.
\label{A10}
\end{equation}

 

\end{document}